\documentclass[sigconf]{acmart}
\AtBeginDocument{%
  }

\setcopyright{acmlicensed}
\copyrightyear{2025}
\acmYear{2025}
\acmDOI{XXXXXXX.XXXXXXX}
\acmConference[CHI '25]{Proceedings of the CHI Conference on Human Factors in Computing Systems}{April 26-- May 1, 2025}{Yokohama, Japan}
\acmBooktitle{CHI 25 Workshop on Mobile Technology and Teens: Understanding the Changing Needs of Sociocultural and Technical Landscapes}
\acmISBN{978-1-4503-XXXX-X/18/06}

\begin{document}

%% The "title" command has an optional parameter,
%% allowing the author to define a "short title" to be used in page headers.
\title{Tao-Technology for Teen Mobile Use: Harmonizing Adaptation, Autonomy, and Reflection}
%%
%% The "author" command and its associated commands are used to define
%% the authors and their affiliations.
%% Of note is the shared affiliation of the first two authors, and the
%% "authornote" and "authornotemark" commands
%% used to denote shared contribution to the research.
\author{Pengyu Zhu}
\orcid{0009-0008-1864-7589}
\affiliation{%
  \institution{National University of Singapore}
  \department{Division of Industrial Design}
  \city{Singapore}
  \country{Singapore}
}
\email{pengyuzhu@u.nus.edu}
\author{Janghee Cho}
\orcid{0000-0002-3193-2180}
\affiliation{%
  \institution{National University of Singapore}
  \department{Division of Industrial Design}
  \city{Singapore}
  \country{Singapore}
}
\email{jcho@nus.edu.sg}

%%
%% By default, the full list of authors will be used in the page
%% headers. Often, this list is too long, and will overlap
%% other information printed in the page headers. This command allows
%% the author to define a more concise list
%% of authors' names for this purpose.
\renewcommand{\shortauthors}{Zhu and Cho}

\begin{teaserfigure}
    \centering
    \includegraphics[width=\textwidth]{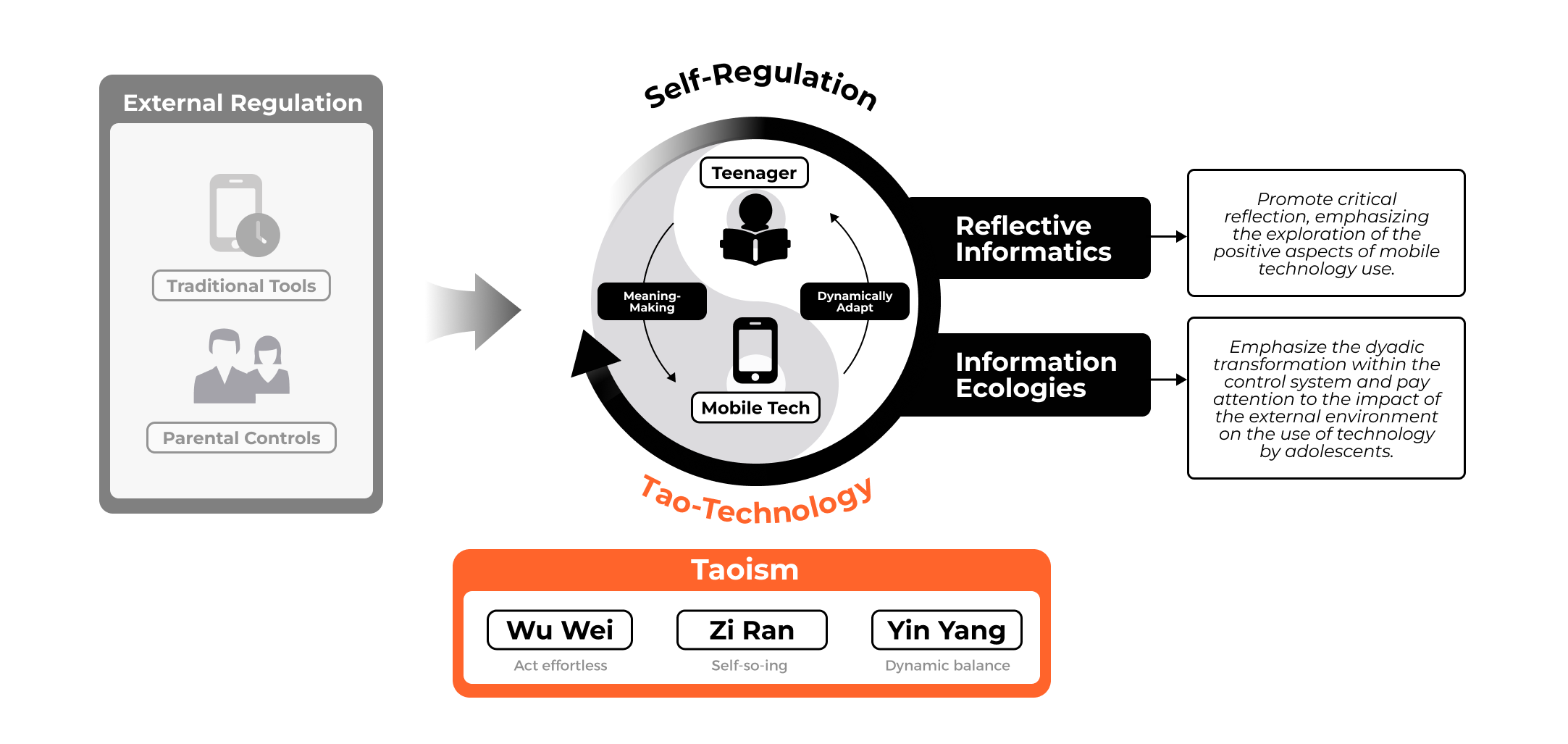} 
    \caption{\textbf{The conceptualized Tao-Technology framework presented in this position paper}}
    \label{fig:sample}
\end{teaserfigure}

%%
%% The abstract is a short summary of the work to be presented in the
%% article.

\begin{abstract}
 
 Adolescents' mobile technology use is often regulated through rigid control mechanisms that fail to account for their autonomy and natural usage patterns. Drawing on Taoist philosophy—particularly Wu Wei, Yin-Yang, and Zi Ran—this position paper proposes Tao-Technology, a self-organizing, adaptive regulatory framework. Integrating insights from Reflective Informatics and Information Ecologies, we explore how mobile technology can dynamically adjust to context while fostering self-reflection and meaning-making. This approach shifts from external restrictions to dynamic co-adaptative regulation, ensuring technology governance remains flexible yet structured, supporting adolescents in cultivating a balanced and intentional relationship with digital technology.

\end{abstract}

%%
%% The code below is generated by the tool at http://dl.acm.org/ccs.cfm.
%% Please copy and paste the code instead of the example below.
%%

%%
%% Keywords. The author(s) should pick words that accurately describe
%% the work being presented. Separate the keywords with commas.
%% A "teaser" image appears between the author and affiliation
%% information and the body of the document, and typically spans the
%% page.

%\received{20 February 2007}
%\received[revised]{12 March 2009}
%\received[accepted]{5 June 2009}

%%
%% This command processes the author and affiliation and title
%% information and builds the first part of the formatted document.
\maketitle

\section{Limitations of Existing Approaches to Controlling Teenagers' Mobile Technology Use}
Teenagers are often described as "mobile natives" \cite{prensky_2001}, having been immersed in smartphone use from an early age and seamlessly integrating mobile technology into their daily lives. However, excessive reliance on mobile devices has been shown to negatively impact their cognitive development and physical well-being \cite{toombs_mushquash2022, sherer_levounis_2022, sanchez_chamorro_my_2024}. Previous research has examined tensions between parental mediation and adolescents' autonomy in use of mobile technology \cite{Chua_What,Lindsey_Techtension, Davis_Relationships}, with a predominant focusing on regulatory strategies. These approaches include (1) negotiation-based parental mediation to balance parental involvement and teenagers' autonomy~\cite{Ananta_Codesign}, (2) social translucence-based regulation to mediate information sharing and privacy in managing parental visibility of information on teens' mobile devices \cite{yardi_social_2011}, and (3) traditional screen-time management strategies~\cite{kawas_when_2021}. However, such strategies do not fully account for parents' varying levels of technology acceptance and may provoke psychological resistance among adolescents due to privacy and autonomy concerns~\cite{MongeMetaReview2023}. Additionally, prior studies tend to overlook how adolescents develop alternative or evasive strategies to circumvent parent controls, thereby undermining their effectiveness~\cite{yardi_social_2011}. 

Although research in the HCI field has increasingly focused on digital self-regulation, relatively little attention has been given to how teenagers themselves develop strategies for managing their mobile technology use \cite{Roffarello_SelfControl}. Studies suggest that adolescents struggle with self-regulation more than adults \cite{VANDEURSEN2015411}, often failing to manage their device use in a deliberate and reflective manner \cite{davis_2023,weinstein_james_2022}. Given the gaps, and the complex socio-technical dynamics in which adolescent interact with mobile technologies at home and school~\cite{magee_four_2017}, it is crucial to move beyond mere external regulation and foster adolescents' awareness and agency in self-regulated mobile device usage. 

Moreover, the boundaries between technology and humans are increasingly fuzzy. As members of Gen-Z and Gen-A, who have grown up alongside mobile technology, adolescents experience its profound influence on their cognitive and behavioral development~\cite{benvenuti_wright_naslund_miers_2023}. This necessitates a shift in how we conceptualize their engagement with mobile technology—not merely as a smart artificial tool for productivity and relaxation but as an extension of the self, a 'digitalized self' co-shaped by self-awareness and technology~\cite{chan_2022}.  At the same time, this raises deeper ontological questions: To what extent does technology shape adolescents, and how do they, in turn, shape their interactions with it?  

This evolving relationship between adolescents and mobile technology signals a need for HCI to move beyond  the traditional  human-centered paradigm \cite{Frauenberger_EntanglementHCI}. As mobile technologies become increasingly entangled with adolescents' identities and ways of experiencing the world, HCI researchers and designers must critically examine not only how technology influences behavior but also how it co-constructs adolescents' sense of self and reality. This leads to a critical questions:  \emph{how can we support adolescents in constructing a reflective understanding of their mobile technology use to cultivate resilient self-regulation strategies?} 

Most existing digital self-control tools rely on external, short-term constraints rather than fostering long-term, intrinsic self-regulation~\cite{lyngs_self-control_2019}. These approaches often lack sustained user motivation and a deeper understanding of the personal impact of mobile technology \cite{MongeMetaReview2023}. Davis et al. developed Locus, a self-regulatory tool designed to facilitate adolescents' reflection on unconscious social media use \cite{Davis2023}. However, its ineffectiveness among some study participants reinforces Lyngs et al.'s argument that externally structured and mechanized reflective guidance through mobile technology is unsustainable \cite{lyngs_self-control_2019}. Beyond external controls, fostering conscious meaning-making in mobile technology use is essential, empowering adolescents to engage with technology in a more intentional and self-aware manner.

In this position paper, we draw on Taoist philosophical perspective to reinterpret self-regulation strategies for adolescent mobile technology use. Specifically we apply the principles of  Wu Wei (effortless action) and Yin Yang \cite{laozi_lau_2003}, which encourage fluid, adaptive self-regulation rather than rigid external control. We believe this approach fosters self-reflection and helps adolescents develop a natural rhythm of engagement with mobile technology. Furthermore, we integrate insights from reflective informatics \cite{baumer_reflective_2015} and information ecologies~\cite{nardi_o’day_1999d} to re-imagine alternative approaches to teenager technology use. %and spiritual practices 
\section{A Tao-Technology: New Rhythms and Patterns of Teenager Technology Use}

\subsection{From Intervention to Adaptation}
In the Taoist classic Tao Te Ching, Lao Tzu introduces the concept of Wu Wei (effortless action), emphasizing alignment with natural rhythms, minimizing excessive intervention, understanding the deepest desires of the heart, and ridding oneself of external disturbances\cite[p. 60]{laozi_lau_2003} in order to achieve a state of harmony between the individual and society. In this state, an individual's behavior is fully consistent with the requirements of the current environment and situation to achieve a state of self-so-ing (Zi Ran) \cite[p.80, p. 82]{laozi_lau_2003}.

Building on this, Allen’s concept of \textit{dao-engineering} provides a theoretical foundation for envisioning \textit{Tao-Technology}--a model of technology use that facilitates to Wu Wei and Zi Ran~\cite{allen_dao_2010}. His work explores how ancient Chinese technologies and artifacts were not rigid control systems but dynamic, evolving entities integrated into their environments~\cite{2007ZhuangZi}. He advocates for a design approach that achieves maximum effect with minimal intervention~\cite[p. 60]{laozi_lau_2003}, so that the development of technology should be dynamic and responsive, seamlessly integrated into the external environment deployed. Inspired by this, Tao-Technology seeks to move away from overly rigid interventions toward a more adaptive and responsive approach to adolescent mobile technology regulation. 

Rather than treating mobile technology as a system that merely reacts to adolescent behavior, Tao-Technology acknowledges adolescents' individual agency and autonomy in regulating their own digital interactions. It operates as a self-organizing regulatory mechanism, dynamically adjusting its strategies in response to evolving usage patterns and cognitive development.  Similar to Kim et al.'s approach, in which machine learning algorithms adapt digital device use interventions based on contextual factors to reinforce positive technology habits \cite{Kim2017}, \textit{Tao-Technology} aims to be non-intrusive, promoting subtle guidance while reinforcing positive technology habits. 

However, while this perspective promotes a less intrusive and more adaptive model of governance, it also raises critical questions about how to balance flexibility with necessary guidance --ensuring that adaptation does not lead to a lack of meaningful structure or accountability. Reflective Informatics~\cite{baumer_reflective_2015} provides an additional layer to this approach, suggesting that strategic reflection-driven interventions could replace traditional rigid control mechanisms. This perspective encourages adolescents not only to critically reassess their technology habits but also to explore personal meaning in their digital engagement. 

\subsection{Dynamic Adaption in A Yin-Yang Perspective}
In Taoist philosophy, the concept of Yin-Yang \cite[p. 103]{laozi_lau_2003} transcends mere binary opposition and instead represents a relational and cyclical interdependence. Within the conceptual framework of \textit{Tao-Technology}, this interplay manifests in the balance between control and autonomy, structure and flexibility, intervention and adaptation. Regulatory mechanisms should not impose rigid rules but instead continuously evolve in response to external factors—mirroring the dynamic balance of Yin-Yang.

Previous research on information ecologies \cite{nardi_o’day_1999d} has identified various factors influencing teenagers' technology use, including local policies and access, affective factors, life stage and goals, and social relationships \cite{magee_four_2017}. This research highlights the interplay and interdependence between individual agency, external conditions, and systemic influences in mobile technology regulation, aligning with the Taoist concept of Yin-Yang \cite[p. 103]{laozi_lau_2003}. Futhermore, it emphasizes that the regulatory mechanisms should continuously integrate inputs from external environmental shifts to account for their impact on adolescents' autonomy. For example, reflective guides could enable teenage users to document relevant emotions and evolving personal goals, ensuring that governance strategies dynamically adjust to maintain an appropriate balance between guidance and self-regulation.

We do not propose Tao-Technology as a fully developed framework but rather as an ongoing conceptual exploration of how Taoist principles might inform adolescent technology regulation. Our aim is to move beyond one-directional control (e.g., screen time restrictions) toward self-reflective regulation, where adolescents actively engage with their technology use rather than passively conforming to external constraints. To advance this discussion, we pose two key questions: (1) How can Taoist principles be systematically integrated into Tao-Technology to inform the design of adaptive mobile technology regulation models? (2) How can reflective interactions within these technological systems support unconscious reflection that leads to conscious self-meaning-making? % Finally, envision how we can support a “spiritual space” for adolescents to reflect on the relationship between self and technology in the midst of such fluid technological rhythms?

\section{Conclusion and Workshop Goals}

We adopt the Taoist concept of Wu Wei, Yin Yang and Zi Ran alongside existing HCI theories to explore an alternative paradigm for mobile technology design in adolescence. This exploration responds to a gap in current digital self-regulation research, which primarily focuses on self-tracking timers and restrictive interventions (lockout mechanisms) \cite{MongeMetaReview2023}. Rather than imposing external constraints, Tao-Technology proposes a co-adaptive relationship between adolescents and mobile technology—encouraging intentional and reflective technology use. By prioritizing reflection over control, this conceptual framework shifts away from mechanistic self-regulation and toward sustained meaning-making in digital habits, that could lead to sustainable behavior change~\cite{Sanches2019}.  Tao-Technology envisions a flexible yet structured model that harmonizes adaptation, autonomy, and reflection, offering a new perspective on adolescent digital well-being.

Our aims for attending this year's workshop are: (1) to provoke a broader discussion on the integration of Taoist Philosophy into teenagers' mobile technology self-control strategies; (2) to explore how reflective design can foster conscious self-meaning-making in adolescent technology use.

\section{Authors' Background in this Research Area}

\textbf{Pengyu Zhu} is a master's student in the Division of Industrial Design at the National University of Singapore. His research explores the ethical dimensions of emerging technology through design research methods, aiming to shape more inclusive technology that empowers people—especially marginalized groups—to reflect on the positive significance of technology use and engage in self-meaning-making.

\textbf{Janghee Cho} is an Assistant Professor in the Division of Industrial Design at National University of Singapore. His research explores
well-being, health, the future of work, and reflective design to
understand how technology can promote sustainable living and
address uncertainty in everyday life. Through a sociotechnical lens,
he examines how technology can foster inclusivity, flourishing, and
meaningful reflective practices.

%%
%% The acknowledgments section is defined using the "acks" environment
%% (and NOT an unnumbered section). This ensures the proper
%% identification of the section in the article metadata, and the
%% consistent spelling of the heading.
%\begin{acks}
%\end{acks}

%%
%% The next two lines define the bibliography style to be used, and
%% the bibliography file.
\bibliographystyle{ACM-Reference-Format}
\bibliography{CHIWORKSHOP}

%%% -*-BibTeX-*-
%%% Do NOT edit. File created by BibTeX with style
%%% ACM-Reference-Format-Journals [18-Jan-2012].

\begin{thebibliography}{28}

%%% ====================================================================
%%% NOTE TO THE USER: you can override these defaults by providing
%%% customized versions of any of these macros before the \bibliography
%%% command.  Each of them MUST provide its own final punctuation,
%%% except for \shownote{}, \showDOI{}, and \showURL{}.  The latter two
%%% do not use final punctuation, in order to avoid confusing it with
%%% the Web address.
%%%
%%% To suppress output of a particular field, define its macro to expand
%%% to an empty string, or better, \unskip, like this:
%%%
%%% \newcommand{\showDOI}[1]{\unskip}   % LaTeX syntax
%%%
%%% \def \showDOI #1{\unskip}           % plain TeX syntax
%%%
%%% ====================================================================

\ifx \showCODEN    \undefined \def \showCODEN     #1{\unskip}     \fi
\ifx \showDOI      \undefined \def \showDOI       #1{#1}\fi
\ifx \showISBNx    \undefined \def \showISBNx     #1{\unskip}     \fi
\ifx \showISBNxiii \undefined \def \showISBNxiii  #1{\unskip}     \fi
\ifx \showISSN     \undefined \def \showISSN      #1{\unskip}     \fi
\ifx \showLCCN     \undefined \def \showLCCN      #1{\unskip}     \fi
\ifx \shownote     \undefined \def \shownote      #1{#1}          \fi
\ifx \showarticletitle \undefined \def \showarticletitle #1{#1}   \fi
\ifx \showURL      \undefined \def \showURL       {\relax}        \fi
% The following commands are used for tagged output and should be
% invisible to TeX
\providecommand\bibfield[2]{#2}
\providecommand\bibinfo[2]{#2}
\providecommand\natexlab[1]{#1}
\providecommand\showeprint[2][]{arXiv:#2}

\bibitem[200(2007)]%
        {2007ZhuangZi}
 \bibinfo{year}{2007}\natexlab{}.
\newblock \bibinfo{booktitle}{\emph{Zhuang Zi}}.
\newblock \bibinfo{publisher}{Shanghai Ancient Books Publishing House}.
\newblock
\showISBNx{9787532546558}
\urldef\tempurl%
\url{https://books.google.com.sg/books?id=MP3FJAAACAAJ}
\showURL{%
\tempurl}


\bibitem[Allen({[n.\,d.]})]%
        {allen_dao_2010}
\bibfield{author}{\bibinfo{person}{Barry Allen}.} \bibinfo{year}{[n.\,d.]}\natexlab{}.
\newblock \showarticletitle{A Dao of Technology?}
\newblock  \bibinfo{volume}{9}, \bibinfo{number}{2} (\bibinfo{year}{[n.\,d.]}), \bibinfo{pages}{151--160}.
\newblock
\showISSN{1540-3009, 1569-7274}
\urldef\tempurl%
\url{https://doi.org/10.1007/s11712-010-9158-1}
\showDOI{\tempurl}


\bibitem[Baumer({[n.\,d.]})]%
        {baumer_reflective_2015}
\bibfield{author}{\bibinfo{person}{Eric~P.S. Baumer}.} \bibinfo{year}{[n.\,d.]}\natexlab{}.
\newblock \showarticletitle{Reflective Informatics: Conceptual Dimensions for Designing Technologies of Reflection}. In \bibinfo{booktitle}{\emph{Proceedings of the 33rd Annual {ACM} Conference on Human Factors in Computing Systems}} (New York, {NY}, {USA}, 2015-04-18) \emph{(\bibinfo{series}{{CHI} '15})}. \bibinfo{publisher}{Association for Computing Machinery}, \bibinfo{pages}{585--594}.
\newblock
\showISBNx{978-1-4503-3145-6}
\urldef\tempurl%
\url{https://doi.org/10.1145/2702123.2702234}
\showDOI{\tempurl}


\bibitem[Benvenuti et~al\mbox{.}(2023)]%
        {benvenuti_wright_naslund_miers_2023}
\bibfield{author}{\bibinfo{person}{Martina Benvenuti}, \bibinfo{person}{Michelle Wright}, \bibinfo{person}{John Naslund}, {and} \bibinfo{person}{Anne~C. Miers}.} \bibinfo{year}{2023}\natexlab{}.
\newblock \showarticletitle{How Technology Use Is Changing Adolescents’ Behaviors and Their social, physical, and Cognitive Development}.
\newblock \bibinfo{journal}{\emph{Current Psychology}}  \bibinfo{volume}{42} (\bibinfo{date}{Jan} \bibinfo{year}{2023}), \bibinfo{pages}{16466–16469}.
\newblock
\urldef\tempurl%
\url{https://doi.org/10.1007/s12144-023-04254-4}
\showDOI{\tempurl}


\bibitem[Blackwell et~al\mbox{.}(2016)]%
        {Lindsey_Techtension}
\bibfield{author}{\bibinfo{person}{Lindsay Blackwell}, \bibinfo{person}{Emma Gardiner}, {and} \bibinfo{person}{Sarita Schoenebeck}.} \bibinfo{year}{2016}\natexlab{}.
\newblock \showarticletitle{Managing Expectations: Technology Tensions among Parents and Teens}. In \bibinfo{booktitle}{\emph{Proceedings of the 19th ACM Conference on Computer-Supported Cooperative Work \& Social Computing}} (San Francisco, California, USA) \emph{(\bibinfo{series}{CSCW '16})}. \bibinfo{publisher}{Association for Computing Machinery}, \bibinfo{address}{New York, NY, USA}, \bibinfo{pages}{1390–1401}.
\newblock
\showISBNx{9781450335928}
\urldef\tempurl%
\url{https://doi.org/10.1145/2818048.2819928}
\showDOI{\tempurl}


\bibitem[Chan(2022)]%
        {chan_2022}
\bibfield{author}{\bibinfo{person}{Kai~Tai Chan}.} \bibinfo{year}{2022}\natexlab{}.
\newblock \showarticletitle{Emergence of the “Digitalized Self” in the Age of Digitalization}.
\newblock \bibinfo{journal}{\emph{Computers in Human Behavior Reports}} \bibinfo{volume}{6}, \bibinfo{number}{2451-9588} (\bibinfo{date}{May} \bibinfo{year}{2022}), \bibinfo{pages}{100191}.
\newblock
\urldef\tempurl%
\url{https://doi.org/10.1016/j.chbr.2022.100191}
\showDOI{\tempurl}


\bibitem[Chowdhury and Bunt(2023)]%
        {Ananta_Codesign}
\bibfield{author}{\bibinfo{person}{Ananta Chowdhury} {and} \bibinfo{person}{Andrea Bunt}.} \bibinfo{year}{2023}\natexlab{}.
\newblock \showarticletitle{Co-Designing with Early Adolescents: Understanding Perceptions of and Design Considerations for Tech-Based Mediation Strategies that Promote Technology Disengagement}. In \bibinfo{booktitle}{\emph{Proceedings of the 2023 CHI Conference on Human Factors in Computing Systems}} (Hamburg, Germany) \emph{(\bibinfo{series}{CHI '23})}. \bibinfo{publisher}{Association for Computing Machinery}, \bibinfo{address}{New York, NY, USA}, Article \bibinfo{articleno}{198}, \bibinfo{numpages}{16}~pages.
\newblock
\showISBNx{9781450394215}
\urldef\tempurl%
\url{https://doi.org/10.1145/3544548.3581134}
\showDOI{\tempurl}


\bibitem[Davis(2023)]%
        {davis_2023}
\bibfield{author}{\bibinfo{person}{Katie Davis}.} \bibinfo{year}{2023}\natexlab{}.
\newblock \bibinfo{booktitle}{\emph{Technology’s Child: Digital Media’s Role in the Ages and Stages of Growing Up.}}
\newblock \bibinfo{publisher}{The MIT Press}.
\newblock
\urldef\tempurl%
\url{https://doi.org/10.7551/mitpress/13406.001.0001}
\showDOI{\tempurl}


\bibitem[Davis et~al\mbox{.}(2019)]%
        {Davis_Relationships}
\bibfield{author}{\bibinfo{person}{Katie Davis}, \bibinfo{person}{Anja Dinhopl}, {and} \bibinfo{person}{Alexis Hiniker}.} \bibinfo{year}{2019}\natexlab{}.
\newblock \showarticletitle{"Everything's the Phone": Understanding the Phone's Supercharged Role in Parent-Teen Relationships}. In \bibinfo{booktitle}{\emph{Proceedings of the 2019 CHI Conference on Human Factors in Computing Systems}} (Glasgow, Scotland Uk) \emph{(\bibinfo{series}{CHI '19})}. \bibinfo{publisher}{Association for Computing Machinery}, \bibinfo{address}{New York, NY, USA}, \bibinfo{pages}{1–14}.
\newblock
\showISBNx{9781450359702}
\urldef\tempurl%
\url{https://doi.org/10.1145/3290605.3300457}
\showDOI{\tempurl}


\bibitem[Davis et~al\mbox{.}(2023)]%
        {Davis2023}
\bibfield{author}{\bibinfo{person}{Katie Davis}, \bibinfo{person}{Petr Slovak}, \bibinfo{person}{Rotem Landesman}, \bibinfo{person}{Caroline Pitt}, \bibinfo{person}{Abdullatif Ghajar}, \bibinfo{person}{Jessica~Lee Schleider}, \bibinfo{person}{Saba Kawas}, \bibinfo{person}{Andrea~Guadalupe Perez~Portillo}, {and} \bibinfo{person}{Nicole~S. Kuhn}.} \bibinfo{year}{2023}\natexlab{}.
\newblock \showarticletitle{Supporting Teens’ Intentional Social Media Use Through Interaction Design: An exploratory proof-of-concept study}. In \bibinfo{booktitle}{\emph{Proceedings of the 22nd Annual ACM Interaction Design and Children Conference}} (Chicago, IL, USA) \emph{(\bibinfo{series}{IDC '23})}. \bibinfo{publisher}{Association for Computing Machinery}, \bibinfo{address}{New York, NY, USA}, \bibinfo{pages}{322–334}.
\newblock
\showISBNx{9798400701313}
\urldef\tempurl%
\url{https://doi.org/10.1145/3585088.3589387}
\showDOI{\tempurl}


\bibitem[Frauenberger(2019)]%
        {Frauenberger_EntanglementHCI}
\bibfield{author}{\bibinfo{person}{Christopher Frauenberger}.} \bibinfo{year}{2019}\natexlab{}.
\newblock \showarticletitle{Entanglement HCI The Next Wave?}
\newblock \bibinfo{journal}{\emph{ACM Trans. Comput.-Hum. Interact.}} \bibinfo{volume}{27}, \bibinfo{number}{1}, Article \bibinfo{articleno}{2} (\bibinfo{date}{Nov.} \bibinfo{year}{2019}), \bibinfo{numpages}{27}~pages.
\newblock
\showISSN{1073-0516}
\urldef\tempurl%
\url{https://doi.org/10.1145/3364998}
\showDOI{\tempurl}


\bibitem[K.~Chua and Mazmanian(2021)]%
        {Chua_What}
\bibfield{author}{\bibinfo{person}{Phoebe K.~Chua} {and} \bibinfo{person}{Melissa Mazmanian}.} \bibinfo{year}{2021}\natexlab{}.
\newblock \showarticletitle{What Are You Doing With Your Phone? How Social Class Frames Parent-Teen Tensions around Teens’ Smartphone Use}. In \bibinfo{booktitle}{\emph{Proceedings of the 2021 CHI Conference on Human Factors in Computing Systems}} (Yokohama, Japan) \emph{(\bibinfo{series}{CHI '21})}. \bibinfo{publisher}{Association for Computing Machinery}, \bibinfo{address}{New York, NY, USA}, Article \bibinfo{articleno}{353}, \bibinfo{numpages}{12}~pages.
\newblock
\showISBNx{9781450380966}
\urldef\tempurl%
\url{https://doi.org/10.1145/3411764.3445275}
\showDOI{\tempurl}


\bibitem[Kawas et~al\mbox{.}(2021)]%
        {kawas_when_2021}
\bibfield{author}{\bibinfo{person}{Saba Kawas}, \bibinfo{person}{Nicole~S. Kuhn}, \bibinfo{person}{Kyle Sorstokke}, \bibinfo{person}{Emily Bascom}, \bibinfo{person}{Alexis Hiniker}, {and} \bibinfo{person}{Katie Davis}.} \bibinfo{year}{2021}\natexlab{}.
\newblock \showarticletitle{When Screen Time Isn't Screen Time: Tensions and Needs Between Tweens and Their Parents During Nature-Based Exploration}. In \bibinfo{booktitle}{\emph{Proceedings of the 2021 {CHI} Conference on Human Factors in Computing Systems}} (New York, {NY}, {USA}, 2021-05-07) \emph{(\bibinfo{series}{{CHI} '21})}. \bibinfo{publisher}{Association for Computing Machinery}, \bibinfo{pages}{1--14}.
\newblock
\showISBNx{978-1-4503-8096-6}
\urldef\tempurl%
\url{https://doi.org/10.1145/3411764.3445142}
\showDOI{\tempurl}


\bibitem[Kim et~al\mbox{.}(2017)]%
        {Kim2017}
\bibfield{author}{\bibinfo{person}{Jaejeung Kim}, \bibinfo{person}{Chiwoo Cho}, {and} \bibinfo{person}{Uichin Lee}.} \bibinfo{year}{2017}\natexlab{}.
\newblock \showarticletitle{Technology Supported Behavior Restriction for Mitigating Self-Interruptions in Multi-device Environments}.
\newblock \bibinfo{journal}{\emph{Proc. ACM Interact. Mob. Wearable Ubiquitous Technol.}} \bibinfo{volume}{1}, \bibinfo{number}{3}, Article \bibinfo{articleno}{64} (\bibinfo{date}{Sept.} \bibinfo{year}{2017}), \bibinfo{numpages}{21}~pages.
\newblock
\urldef\tempurl%
\url{https://doi.org/10.1145/3130932}
\showDOI{\tempurl}


\bibitem[Laozi and Lau(2003)]%
        {laozi_lau_2003}
\bibfield{author}{\bibinfo{person}{Laozi} {and} \bibinfo{person}{D~C Lau}.} \bibinfo{year}{2003}\natexlab{}.
\newblock \bibinfo{booktitle}{\emph{Tao te ching}}.
\newblock \bibinfo{publisher}{Penguin}, \bibinfo{address}{London}.
\newblock
\showISBNx{9780140441314}


\bibitem[Lyngs et~al\mbox{.}(2019)]%
        {lyngs_self-control_2019}
\bibfield{author}{\bibinfo{person}{Ulrik Lyngs}, \bibinfo{person}{Kai Lukoff}, \bibinfo{person}{Petr Slovak}, \bibinfo{person}{Reuben Binns}, \bibinfo{person}{Adam Slack}, \bibinfo{person}{Michael Inzlicht}, \bibinfo{person}{Max Van~Kleek}, {and} \bibinfo{person}{Nigel Shadbolt}.} \bibinfo{year}{2019}\natexlab{}.
\newblock \showarticletitle{Self-Control in Cyberspace: Applying Dual Systems Theory to a Review of Digital Self-Control Tools}. In \bibinfo{booktitle}{\emph{Proceedings of the 2019 {CHI} Conference on Human Factors in Computing Systems}} (Glasgow Scotland Uk, 2019-05-02). \bibinfo{publisher}{{ACM}}, \bibinfo{pages}{1--18}.
\newblock
\showISBNx{978-1-4503-5970-2}
\urldef\tempurl%
\url{https://doi.org/10.1145/3290605.3300361}
\showDOI{\tempurl}


\bibitem[Magee et~al\mbox{.}(2017)]%
        {magee_four_2017}
\bibfield{author}{\bibinfo{person}{Rachel~M. Magee}, \bibinfo{person}{Denise~E. Agosto}, {and} \bibinfo{person}{Andrea Forte}.} \bibinfo{year}{2017}\natexlab{}.
\newblock \showarticletitle{Four Factors that Regulate Teen Technology Use in Everyday Life}. In \bibinfo{booktitle}{\emph{Proceedings of the 2017 {ACM} Conference on Computer Supported Cooperative Work and Social Computing}} (New York, {NY}, {USA}, 2017-02-25) \emph{(\bibinfo{series}{{CSCW} '17})}. \bibinfo{publisher}{Association for Computing Machinery}, \bibinfo{pages}{511--522}.
\newblock
\showISBNx{978-1-4503-4335-0}
\urldef\tempurl%
\url{https://doi.org/10.1145/2998181.2998310}
\showDOI{\tempurl}


\bibitem[Nardi and O’day(1999)]%
        {nardi_o’day_1999d}
\bibfield{author}{\bibinfo{person}{Bonnie~A Nardi} {and} \bibinfo{person}{Vicki O’day}.} \bibinfo{year}{1999}\natexlab{}.
\newblock \bibinfo{booktitle}{\emph{Information ecologies : Using technology with heart}}.
\newblock \bibinfo{publisher}{Mit Press}, \bibinfo{address}{Cambridge, Massachusetts}.
\newblock
\showISBNx{9780262140669}


\bibitem[Prensky(2001)]%
        {prensky_2001}
\bibfield{author}{\bibinfo{person}{Marc Prensky}.} \bibinfo{year}{2001}\natexlab{}.
\newblock \showarticletitle{Digital Natives, Digital Immigrants Part 2: Do They Really Think Differently?}
\newblock \bibinfo{journal}{\emph{On the Horizon}} \bibinfo{volume}{9}, \bibinfo{number}{6} (\bibinfo{date}{Nov} \bibinfo{year}{2001}), \bibinfo{pages}{1–6}.
\newblock
\urldef\tempurl%
\url{https://doi.org/10.1108/10748120110424843}
\showDOI{\tempurl}


\bibitem[Roffarello and De~Russis(2023a)]%
        {MongeMetaReview2023}
\bibfield{author}{\bibinfo{person}{Alberto~Monge Roffarello} {and} \bibinfo{person}{Luigi De~Russis}.} \bibinfo{year}{2023}\natexlab{a}.
\newblock \showarticletitle{Achieving Digital Wellbeing Through Digital Self-control Tools: A Systematic Review and Meta-analysis}.
\newblock \bibinfo{journal}{\emph{ACM Trans. Comput.-Hum. Interact.}} \bibinfo{volume}{30}, \bibinfo{number}{4}, Article \bibinfo{articleno}{53} (\bibinfo{date}{Sept.} \bibinfo{year}{2023}), \bibinfo{numpages}{66}~pages.
\newblock
\showISSN{1073-0516}
\urldef\tempurl%
\url{https://doi.org/10.1145/3571810}
\showDOI{\tempurl}


\bibitem[Roffarello and De~Russis(2023b)]%
        {Roffarello_SelfControl}
\bibfield{author}{\bibinfo{person}{Alberto~Monge Roffarello} {and} \bibinfo{person}{Luigi De~Russis}.} \bibinfo{year}{2023}\natexlab{b}.
\newblock \showarticletitle{Achieving Digital Wellbeing Through Digital Self-control Tools: A Systematic Review and Meta-analysis}.
\newblock \bibinfo{journal}{\emph{ACM Trans. Comput.-Hum. Interact.}} \bibinfo{volume}{30}, \bibinfo{number}{4}, Article \bibinfo{articleno}{53} (\bibinfo{date}{Sept.} \bibinfo{year}{2023}), \bibinfo{numpages}{66}~pages.
\newblock
\showISSN{1073-0516}
\urldef\tempurl%
\url{https://doi.org/10.1145/3571810}
\showDOI{\tempurl}


\bibitem[Sanches et~al\mbox{.}(2019)]%
        {Sanches2019}
\bibfield{author}{\bibinfo{person}{Pedro Sanches}, \bibinfo{person}{Axel Janson}, \bibinfo{person}{Pavel Karpashevich}, \bibinfo{person}{Camille Nadal}, \bibinfo{person}{Chengcheng Qu}, \bibinfo{person}{Claudia Daud\'{e}n~Roquet}, \bibinfo{person}{Muhammad Umair}, \bibinfo{person}{Charles Windlin}, \bibinfo{person}{Gavin Doherty}, \bibinfo{person}{Kristina H\"{o}\"{o}k}, {and} \bibinfo{person}{Corina Sas}.} \bibinfo{year}{2019}\natexlab{}.
\newblock \showarticletitle{HCI and Affective Health: Taking stock of a decade of studies and charting future research directions}. In \bibinfo{booktitle}{\emph{Proceedings of the 2019 CHI Conference on Human Factors in Computing Systems}} (Glasgow, Scotland Uk) \emph{(\bibinfo{series}{CHI '19})}. \bibinfo{publisher}{Association for Computing Machinery}, \bibinfo{address}{New York, NY, USA}, \bibinfo{pages}{1–17}.
\newblock
\showISBNx{9781450359702}
\urldef\tempurl%
\url{https://doi.org/10.1145/3290605.3300475}
\showDOI{\tempurl}


\bibitem[Sanchez~Chamorro et~al\mbox{.}(2024)]%
        {sanchez_chamorro_my_2024}
\bibfield{author}{\bibinfo{person}{Lorena Sanchez~Chamorro}, \bibinfo{person}{Carine Lallemand}, {and} \bibinfo{person}{Colin~M. Gray}.} \bibinfo{year}{2024}\natexlab{}.
\newblock \showarticletitle{"My Mother Told Me These Things are Always Fake" - Understanding Teenagers' Experiences with Manipulative Designs}. In \bibinfo{booktitle}{\emph{Designing Interactive Systems Conference}} ({IT} University of Copenhagen Denmark, 2024-07). \bibinfo{publisher}{{ACM}}, \bibinfo{pages}{1469--1482}.
\newblock
\showISBNx{979-8-4007-0583-0}
\urldef\tempurl%
\url{https://doi.org/10.1145/3643834.3660704}
\showDOI{\tempurl}


\bibitem[Sherer and Levounis(2022)]%
        {sherer_levounis_2022}
\bibfield{author}{\bibinfo{person}{James Sherer} {and} \bibinfo{person}{Petros Levounis}.} \bibinfo{year}{2022}\natexlab{}.
\newblock \showarticletitle{Technological Addictions}.
\newblock \bibinfo{journal}{\emph{Current Psychiatry Reports}} \bibinfo{volume}{24}, \bibinfo{number}{9} (\bibinfo{date}{Jul} \bibinfo{year}{2022}), \bibinfo{pages}{399–406}.
\newblock
\urldef\tempurl%
\url{https://doi.org/10.1007/s11920-022-01351-2}
\showDOI{\tempurl}


\bibitem[Toombs et~al\mbox{.}(2022)]%
        {toombs_mushquash2022}
\bibfield{author}{\bibinfo{person}{Elaine Toombs}, \bibinfo{person}{Christopher~J. Mushquash}, \bibinfo{person}{Linda Mah}, \bibinfo{person}{Kathy Short}, \bibinfo{person}{Nancy~L. Young}, \bibinfo{person}{Chiachen Cheng}, \bibinfo{person}{Lynn Zhu}, \bibinfo{person}{Gillian Strudwick}, \bibinfo{person}{Catherine Birken}, \bibinfo{person}{Jessica Hopkins}, \bibinfo{person}{Daphne~J. Korczak}, \bibinfo{person}{Anna Perkhun}, {and} \bibinfo{person}{Karen~B. Born}.} \bibinfo{year}{2022}\natexlab{}.
\newblock \showarticletitle{Increased Screen Time for Children and Youth During the COVID-19 Pandemic}.
\newblock \bibinfo{journal}{\emph{Science Table}} \bibinfo{volume}{3}, \bibinfo{number}{59} (\bibinfo{date}{Apr} \bibinfo{year}{2022}).
\newblock
\urldef\tempurl%
\url{https://doi.org/10.47326/ocsat.2022.03.59.1.0}
\showDOI{\tempurl}


\bibitem[{van Deursen} et~al\mbox{.}(2015)]%
        {VANDEURSEN2015411}
\bibfield{author}{\bibinfo{person}{Alexander~J.A.M. {van Deursen}}, \bibinfo{person}{Colin~L. Bolle}, \bibinfo{person}{Sabrina~M. Hegner}, {and} \bibinfo{person}{Piet~A.M. Kommers}.} \bibinfo{year}{2015}\natexlab{}.
\newblock \showarticletitle{Modeling habitual and addictive smartphone behavior: The role of smartphone usage types, emotional intelligence, social stress, self-regulation, age, and gender}.
\newblock \bibinfo{journal}{\emph{Computers in Human Behavior}}  \bibinfo{volume}{45} (\bibinfo{year}{2015}), \bibinfo{pages}{411--420}.
\newblock
\showISSN{0747-5632}
\urldef\tempurl%
\url{https://doi.org/10.1016/j.chb.2014.12.039}
\showDOI{\tempurl}


\bibitem[Weinstein and James(2022)]%
        {weinstein_james_2022}
\bibfield{author}{\bibinfo{person}{Emily Weinstein} {and} \bibinfo{person}{Carrie James}.} \bibinfo{year}{2022}\natexlab{}.
\newblock \bibinfo{booktitle}{\emph{Behind their screens : what teens are facing (and adults are missing)}}.
\newblock \bibinfo{publisher}{The MIT Press}, \bibinfo{address}{Cambridge, Massachusetts}.
\newblock
\showISBNx{9780262047357}


\bibitem[Yardi and Bruckman(2011)]%
        {yardi_social_2011}
\bibfield{author}{\bibinfo{person}{Sarita Yardi} {and} \bibinfo{person}{Amy Bruckman}.} \bibinfo{year}{2011}\natexlab{}.
\newblock \showarticletitle{Social and technical challenges in parenting teens' social media use}. In \bibinfo{booktitle}{\emph{Proceedings of the {SIGCHI} Conference on Human Factors in Computing Systems}} (New York, {NY}, {USA}, 2011-05-07) \emph{(\bibinfo{series}{{CHI} '11})}. \bibinfo{publisher}{Association for Computing Machinery}, \bibinfo{pages}{3237--3246}.
\newblock
\showISBNx{978-1-4503-0228-9}
\urldef\tempurl%
\url{https://doi.org/10.1145/1978942.1979422}
\showDOI{\tempurl}


\end{thebibliography}

%%
%% If your work has an appendix, this is the place to put it.

\end{document}